\documentclass[12pt,a4paper,dvips]{article}
\usepackage{epsfig,wrapfig,times,mathptm}
\newcommand{\degr}{^\circ}
\renewcommand{\section}[1]{\vspace{6pt} \noindent\mbox{#1} \newline
\noindent}
\renewcommand{\subsection}[1]{\vspace{6pt}
\noindent\mbox{\underline{#1}}
\newline \noindent}
\renewcommand{\subsubsection}[1]{\vspace{6pt}
\noindent\mbox{\underline{#1}}
\noindent}

\newfont{\sansb}{cmssbx10}
\newfont{\sans}{cmss10}

\newcommand\g{{$\gamma$}}
\newcommand\apj{ApJ}

\newcommand{\Cv}{\v{C}erenkov}

\begin{document}
{\small OG 4.1.14 \vspace{-24pt}\\}     
{\center \LARGE VERY HIGH ENERGY OBSERVATIONS \\ OF PSR~B1951+32
\vspace{6pt}\\}
R.Srinivasan$^1$,
 P.J.Boyle$^2$, 
 J.H.Buckley$^3$, 
 A.M.Burdett$^4$, 
 J.Bussons Gordo$^2$, 
 D.A.Carter-Lewis$^5$,
  
 M.F.Cawley$^6$, 
 M.Catanese$^5$, 
 E. Colombo$^8$,  
 D.J.Fegan$^2$, 
 J.P.Finley$^1$, 
 J.A.Gaidos$^1$,
 A.M.Hillas$^4$,
 
 R.C.Lamb$^6$, 
 F.Krennrich$^5$, 
 R.W.Lessard$^1$, 
 C.Masterson$^2$, 
 J.E.McEnery$^2$,
 G.Mohanty$^5$,
 P. Moriarty$^7$, 
 J.Quinn$^2$,
 A.J.Rodgers$^4$, 
 H.J.Rose$^4$, 
 F.W.Samuelson$^5$, 
 G.H.Sembroski$^1$, 
 T.C.Weekes$^3$, 
 and J.Zweerink$^5$ 
\vspace{6pt}\\
{\it $^1$Department of Physics, Purdue University, West
Lafayette, IN 47907 \\
$^2$ Physics Department, University College, Dublin 4,
Ireland \\
$^3$ Whipple Observatory, Harvard-Smithsonian
CfA, P.O. Box 97, Amado, AZ 85645-0097 \\
$^4$ Department of Physics and Astronomy, University of Leeds,
Leeds, LS2 9JT, Yorkshire, UK \\ 
$^5$ Department of Physics and Astronomy, Iowa State University,
Ames, IA 50011-3160  \\ 
$^6$ Space Radiation Lab, California Institute of Technology,
Pasadena, CA 91125  \\ 
$^7$ Regional Technical College, Galway, Ireland  \\ 
$^8$ Present address: CONAE, Paseo Colon 751, Argentina
\vspace{-12pt}\\}

{\center ABSTRACT\\}

PSR B1951+32 is a \g-ray pulsar detected by the {\it{Energetic Gamma Ray
Experiment Telescope}} (EGRET) and identified with the 39.5 ms radio
pulsar in the supernova remnant CTB 80. 
The EGRET data shows no evidence for a spectral turnover. 
Here we report on the first observations
of PSR B1951+32 beyond 30 GeV. The observations were carried out with
the 10m \g-ray telescope at the Whipple Observatory on
Mt. Hopkins, Arizona. In 8.1 hours of observation we find
no evidence for steady or periodic emission from PSR B1951+32
above $\sim$ 260 GeV. FLux upper limits are 
derived and compared with model extrapolations from lower energies and
the predictions of emission models.

\setlength{\parindent}{1cm}
\section{INTRODUCTION}
The pursuit of 
Very High Energy (VHE) astrophysics has resulted in the discovery
of five sources, of which three are associated with  young spin-powered 
pulsars.  
VHE emission has been detected from the direction of the Crab Nebula  
(Vacanti et.al., 1991), 
the Vela pulsar (Takanori 1996) and PSR B1706-44 (Kifune et al., 1995) 
but no evidence has been found for periodic emission at these energies
in these experiments.  

PSR B1951+32 has been detected as a pulsating 
X-ray source (Safi-Harb et al., 1995)
and as a high energy \g-ray pulsar at E $\ge$ 100 MeV
at the radio period (Ramanamurthy et al., 1995).  
It can be inferred from the five pulsars seen in the MeV to GeV
$\gamma$-ray region that longer period or older ( $\sim$ 10$^5$ years ) 
pulsars have a greater fraction of spin down energy emitted as high energy
\g-rays. 
The best fit outer gap
model of Zhang and Cheng (1997) suggests that PSR B1951+32 should emit
detectable levels of TeV \g-rays 
(Figure \ref{fig2}). 
The multiwavelength  spectrum of PSR B1951+32 (Figure 1b)
indicates a maximum power per decade at energies consistent with a few GeV
and still rising at 10 GeV.
These factors make PSR B1951+32 a good
candidate for observations with the ACT above 100 GeV.  

\section{OBSERVATIONS} 
The observations of PSR B1951+32 reported here were acquired 
with the 10m reflector located
at the Whipple Observatory on Mt. Hopkins in Arizona.  
A total of 14 $\it{Tracking}$ runs and 4 $\it{On/Off}$ pairs taken 
between 13th May, 1996 and 17th July, 1996 constitute the database
for all subsequent discussion.  
The total $\it{On}$  source observing time
is 8.1 hrs.  
The radio position (J2000) of PSR B1951+32 ($\alpha$
= 19$^h$ 52$^m$ 58.25$^s$, $\delta$= 32$^{\circ}$ 52$^{\prime}$
40.9$^{\prime\prime}$) was assumed for
the subsequent timing analysis. 
\newpage  
\begin{table}[h]
\vspace{-12pt}
\caption{Pulsar Parameters} \label{tab1}
\begin{center}
\begin{tabular}{lccccc} \hline \hline
PSR  & P & $\dot{P}$ &
Distance  & log$_{10}$B  & Log$_{10}$$\dot{E}$ \\
 & msec & 10$^{-15}$ss$^{-1}$ & kpc & Gauss & ergs/s \\ \hline
B1951+32 & 39.53& 5.8494 & 2.5 & 11.69& 36.57 \\ \hline \hline
\end{tabular}
\end{center}
\end{table}

\section{ANALYSIS AND RESULTS}    
\subsection{Standard Analysis}  
The event selection criteria are collectively called $\it{Supercuts95}$
and a detailed description can be found elsewhere (Catanese et al., 1995).
$\it{Supercuts95}$ raises the 
effective energy threshold of the detector with its software trigger and
$\it{size}$ cuts. 
PSR B1951+32 
appears to have a steep spectrum at EGRET energies and 
since the pulsar spectrum is expected to cut off, 
it behooves us to 
reduce the threshold of our analysis to search for a lower energy signal.  
The dominant background at lower energies is due to muons whose 
images appear in the camera as arcs and can be discriminated by
a cut on their large $\it{length/size}$ values.
Hence the selection criteria used 
$\it{Supercuts95}$ on images with sizes larger than 400 p.e. and 
$\it{Smallcuts}$ (Table \ref{tab2}) for images 
with sizes less than 400 p.e.
\begin{table}[h]
\vspace{-12pt}
\caption{Parameter ranges for selecting \g-ray images}\label{tab2}
\begin{center}
\begin{tabular}{lcc} 
\hline \hline 
Parameter & Supercuts95 & Smallcuts \\ \hline 
length & 0${\degr}$16 - 0${\degr}$30& unchanged\\
width  & 0${\degr}$073 - 0${\degr}$15 & unchanged \\
distance & 0${\degr}51$ - 1${\degr}$1  & unchanged \\
alpha    & $<$ 15$^{\degr}$ & unchanged \\
max1     & $>$ 100 p.e. & 45 p.e. - 100 p.e. \\
max2     & $>$ 80 p.e.  & 45 p.e. -  80 p.e. \\
size     & $\geq$ 400 p.e. & 0 -  400 p.e. \\
length/size &  not used & $<$ 7.5 $\times$ 10$^{-4}$ $^{\circ}$/p.e. \\
\hline \hline 
\end{tabular}
\end{center}
\end{table}
\begin{table}[h]
\vspace{-12pt}
\caption{Selected Events for Steady Emission analysis} \label{tab3}
\begin{center}
\begin{tabular}{lcccc} \hline \hline
Selection&Source Events & Background Events & Excess &Significance \\
 & $\alpha$ $<$ 15$^{\circ}$ &
$\alpha$ $<$ 15$^{\circ}$ &  &   \\ \hline
Supercuts95 &292 & 254  &  38 &   1.16$\sigma$ \\
Smallcuts  &618 & 672& -54& -1.10$\sigma$  \\
Supercuts95 + Smallcuts  & 910  & 926 & -16  & -0.24$\sigma$ \\
\hline
\hline
\end{tabular}
\end{center}
\end{table}
No steady emission is detected from PSR B1951+32 and  
3$\sigma$ flux upper limits 
are displayed in Table \ref{tab5}.
The effective area for $\it{Supercuts95}$, that was used
to calculate the upper limit, was taken as A$_{\it{eff}}$ $\sim$
3.5 $\times$ 10$^8$ cm$^{2}$: the same area was used for the
dataset that resulted from a combination of $\it{Supercuts95}$
and $\it{Smallcuts}$ although here there is more systematic
uncertainty.
The energy threshold
was obtained from simulations and extrapolating 
the Crab Nebula \g-ray rate for each set of cuts used assuming a spectrum
$\sim$ E$^{-2.4}$. 

\subsection{Periodic Analysis}
The arrival times of the \Cv\  events were registered by a GPS
clock with an absolute resolution of 250 $\mu$sec. An oscillator
calibrated by GPS second marks was used to interpolate to a resolution
of 0.1 $\mu$sec.
After an oscillator drift correction was
applied, all arrival times were transformed to the solar system barycenter
and folded to produce the phases, $\phi$$_{j}$, of the
events modulo the pulse period. 
The ephemeris frequency parameters used were
$\nu$= 25.2963719901267 s$^{-1}$ and
$\dot{\nu}$=-3.73940$\times$10$^{-12}$s$^{-2}$, at the epoch
t$_0$=JD 2450177.5.
This frequency was extrapolated 72 days to obtain a timing solution relevant
to the epoch of observation. The datasets, however, were taken within the
validity interval of the above ephemeris. 

To check the Whipple Observatory timing systems an $\underline{optical}$ 
observation of the Crab pulsar was undertaken on December 2nd (UT) 1996
using the 10m reflector. The phase analysis of the event arrival times 
yielded a clear detection of optical Crab pulsar signal which is in
phase with the radio pulse and demonstrates the validity
of the timing, data acquisition and software in the presence of a
pulsed signal. 
No evidence of pulsed emission from PSR B1951+32 
at the radio period exists. To calculate a pulsed flux
upper limit we assumed the same pulse profile as seen at EGRET
energies, i.e. with the phase range for the main
pulse and secondary pulse as 0.12 - 0.22 and 0.48 - 0.74 respectively
(Ramanamurthy et al., 1995). 

\begin{table}[h]
\vspace{-12pt}
\caption{Integral Flux Upper limits} \label{tab5}
\begin{center}
\begin{tabular}{lccc} \hline \hline 
 &Steady Emission & Periodic Emission & Threshold \\
 & cm$^{-2}$s$^{-1}$)
& (cm$^{-2}$s$^{-1}$) & (GeV) \\ \hline \hline
Supercuts95  & 0.97 $\times$ 10$^{-11}$
& 3.7$\times$ 10$^{-12}$ & $\geq$ 370   \\
Supercuts95 + Smallcuts  & 1.95$\times$ 10$^{-11}$
& 6.7$\times$ 10$^{-12}$
& $\geq$ 260  \\
\hline \hline 
\end{tabular}
\end{center}
\end{table}
\begin{wrapfigure}[21]{r}{7.5cm}
\epsfig{file=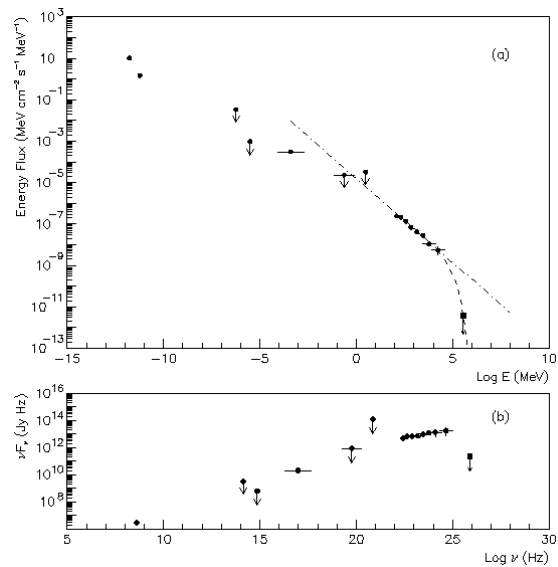,width=7.5cm}
\caption{\it The pulsed energy spectrum of PSR B1951+32.
The Whipple limit is indicated as a filled square at 370 GeV. 
(See text for details). \label{fig1} } 
\end{wrapfigure}
\section{DISCUSSION}
PSR B1951+32 is surrounded by a compact nebula which gives a
plerionic nature to the supernova remnant, CTB80.
X-ray plerions are good candidates for VHE emission since the
electrons responsible for nebular synchrotron X-rays
should also create VHE \g-rays via the inverse Compton (IC) process. 
It is expected
that for plerions, such as that associated with PSR B1951+32 where
the density of nebular synchrotron photons is too low for SSC to take
place, detectable VHE emission should be produced by the IC scattering
of the 2.7K cosmic microwave background by the same electrons radiating 
synchrotron X-ray photons. Interpreting the unpulsed
X-ray emission form CTB80 as the synchrotron emission from a plerion, the
estimated IC flux $>$ 1 TeV is 6.6 $\times$ 10$^{-13}$ TeV/cm$^2$/s/TeV
(De Jager et. al., 1995). This represents the lower limit on the IC flux since
there can be other sources of soft photons in addition to the microwave
background. 

To model the pulsed high energy spectrum, a function of the form 
\begin{equation}
dN_{\gamma}/dE = K E^{-\Gamma}e^{(-E/E_{0})}
\label{eq}
\end{equation}
was used where E is the photon energy, $\Gamma$ is the photon
spectral index and E$_0$ is the cut off energy. 
The pulsed upper limit reported here
is two orders of magnitude lower than the extrapolated EGRET 
power law. Equation (2) was used to extrapolate the EGRET
spectrum to VHE energies constrained by the TeV upper limit reported      
here and indicates a cut off energy of E$_{0}$ $\leq$ 75 GeV for pulsed
emission (Figure 1a).

\begin{wrapfigure}[22]{r}{7.6cm}
\epsfig{file=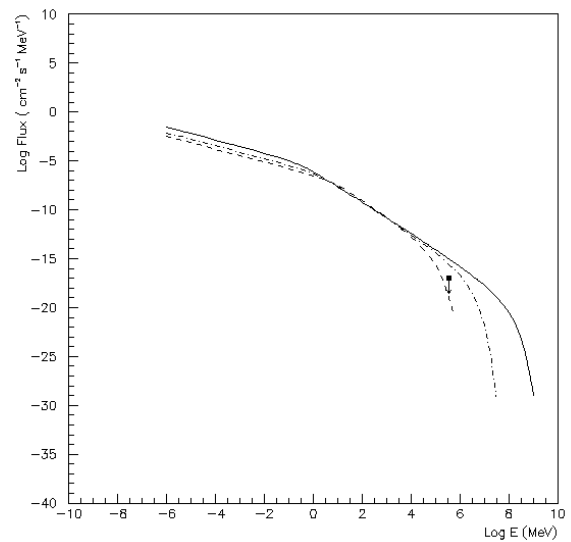,width=7.6cm}
\caption{\it Predicted pulsed \g-ray flux of PSR B1951+32 from
the Zhang and Cheng outergap model. The solid, dot-dash and 
dashed curves correspond to $\alpha$=0.5, 0.6, 0.7 respectively.
(See text for details). \label{fig2}} 
\end{wrapfigure}

The strength of the cut off provides a good discriminant
between the various pulsar emission models.
The status of current observations and the derived cutoff
discussed above indicates that the cutoff is beyond 10 GeV.
In polar cap models this would indicate a sharp cutoff since
the pair production optical depth increases exponentially
with photon energy (Harding 1997).  However, it is not
possible to constrain the shape of the cut off with the non
detection of pulsed TeV flux reported here.
The most relavant comparison of the Whipple upper limit with 
emission models is the outer gap model of Zhang and Cheng 
(see Figure \ref{fig2}). This model includes
the effect of geometry in the treatment of pulsed emission via a parameter
$\alpha$ = r/r$_{L}$, the radial distance to the synchrotron emitting
region near the outer gap, r, as a function of the light cylinder radius
r$_{L}$. Our pulsed upper limits are consistent with the outer gap model
if $\alpha$ $>$ 0.6 implying an emission region far out in the
magnetosphere. 

The result reported here is the first observation of PSR B1951+32
beyond 30 GeV. 
PSR B1951+32 exhibits very similar spectral
behavior and morphological features, such as an associated synchrotron
nebula, to PSR B1706-44 (Finley et al., 1997). If these factors are any
indication
of similar emission mechanisms in pulsars then the lack of unpulsed
emission from PSR B1951+32 is puzzling considering that PSR B1706-44
was detected as a VHE source of unpulsed emission $>$ 1 TeV
(Kifune et al., 1995).
Lack of pulsed emission
indicates that the processes producing pulsed high energy photons over
two decades of energy in the EGRET energy range somehow become
ineffective
over a decade of energy to result in a lack of VHE \g-rays. The low
magnetic field of PSR B1951+32 relative to the average pulsar field
implies
that attenuation of \g-rays by magnetic absorption is not a likely
explanation
for the non-detection.

\section{ACKNOWLEDGEMENTS}
We acknowledge the technical assistance of K. Harris. This research
is supported by grants from the U.S. Department of Energy, NASA,
PPARC in the UK and by Forbairt in Ireland. The authors wish to
thank
A. Lyne for providing the radio ephemeris of PSR B1951+32 and D. J.
Thompson
for providing the multiwavelength spectrum for PSR B1951+32.

\section{REFERENCES} 
\setlength{\parindent}{-5mm}
\begin{list}{}{\topsep 0pt \partopsep 0pt \itemsep 0pt \leftmargin 5mm
\parsep 0pt \itemindent -5mm}
\vspace{-15pt}
\item Catanese, M. et al. 1995, Towards a Major Atmospheric Cherenkov Detector -IV, Padova, 335
\item De Jager, O.C. et al. 1995, 
in Proc. 24th ICRC (Rome), 2, 528
\item Finley, J.P. et al. 1997, \apj, in preparation
\item Kifune, T. et al. 1995, \apj, 438, L91
\item Ramanamurthy, P.V., et al. 1995,
\apj, 447, L109
\item Safi-Harb, S., O\"gelman, H.,
\& Finley, J.P. 1995, \apj, 439, 722
\item Harding, A. 1997, private communication
\item Takanori, Y. 1996, Ph.D. thesis,
University of Tokyo
\item Vacanti, G., 1991, \apj, 377, 467
\item Zhang, L. and Cheng, K.S. 1997, \apj,
submitted
\end{list}

\end{document}